\documentclass[longnamesfirst]{emulateapj}

\shorttitle{FUSE Detection of \ion{O}{6} from NGC\,6543}
\shortauthors{Gruendl et al.}

\newcommand{\fuse}     {{\it FUSE}}

\newcommand{\CIV}      {\ion{C}{4}}

\newcommand{\NV}       {\ion{N}{5}}
\newcommand{\OIII}     {[\ion{O}{3}]}
\newcommand{\OV}       {\ion{O}{5}}
\newcommand{\OVI}      {\ion{O}{6}}

\begin{document}
\title{\fuse\ Observations of Nebular \OVI\ Emission from
NGC\,6543\footnote{Based on observations made with the 
NASA-CNES-CSA Far Ultraviolet Spectroscopic Explorer.
{\it FUSE} is operated for NASA by the 
Johns Hopkins University under NASA contract NAS~5-32985.}}

\author{Robert A.\ Gruendl, You-Hua Chu}
\affil{Astronomy Department, University of Illinois at Urbana-Champaign\\
1002 West Green Street, Urbana, IL 61801}
\email{gruendl@astro.uiuc.edu, chu@astro.uiuc.edu}

\and

\author{Mart\'{\i}n A.\ Guerrero}
\affil{Instituto de Astrof\'{\i}sica de Andaluc\'{\i}a, CSIC\\
Apartado Correos 3004, E-18080, Granada, Spain}
\email{mar@iaa.es}


\received{}
\accepted{}

\begin{abstract}

NGC\,6543 is one of the few planetary nebulae (PNe) whose X-ray
emission has been shown to be extended and originate from hot 
interior gas.
Using \fuse\ observations we have now detected nebular \OVI\ 
emission from NGC\,6543.  
Its central star, with an effective temperature of $\sim$50,000~K,
is too cool to photoionize \OV, so the \OVI\ ions must have been 
produced by thermal collisions at the interface between the hot 
interior gas and the cool nebular shell.
We modeled the \OVI\ emission incorporating thermal conduction, but 
find that simplistic assumptions for the AGB and fast wind mass loss 
rates overproduce X-ray emission and \OVI\ emission.
We have therefore adopted the pressure of the interior hot gas 
for the interface layer and find that expected \OVI\ emission to be
comparable to the observations.

\end{abstract}


\section{Introduction}

Planetary nebulae (PNe) are formed through dynamic interactions
between the current fast stellar wind and previous asymptotic 
giant branch (AGB) wind \citep{K83,Fetal90}; thus, the interior of a 
PN is filled with shocked fast wind.
This hot interior gas and the cool nebular shell 
form a contact discontinuity where heat conduction \citep{S62} 
is expected to occur.
The resulting mass evaporation from the dense nebular shell into 
the hot interior lowers the temperature and raises the density of the 
hot gas \citep{Wetal77}, significantly increasing the X-ray 
emissivity.
Hydrodynamic models of PNe with heat conduction predict X-ray 
emission that should be easily detectable with modern X-ray 
observatories \citep{ZP96}.

{\it Chandra} and {\it XMM-Newton} have indeed detected diffuse X-ray 
emission from PNe, with plasma temperatures of $1-3\times10^6$~K and 
X-ray luminosities of $L_{\rm X}=3-100\times10^{31}$ erg s$^{-1}$ 
\citep{GCG04}.
The limb-brightened X-ray morphology and the low plasma 
temperatures are qualitatively consistent with the predictions of 
models with heat conduction; however, the observed $L_{\rm X}$ are 
generally too low.
For example, the observed $L_{\rm X}$ of NGC\,6543, $1\times10^{32}$
ergs~s$^{-1}$ \citep{Cetal01}, is an order of magnitude lower 
than that modeled by \citet{ZP98}.
Similar discrepancies between observed and modeled $L_{\rm X}$ 
have been seen in Wolf-Rayet bubbles \citep[e.g., NGC\,6888,][]{GLM96} 
and superbubbles \citep[e.g., M17,][]{Detal03}, suggesting that this
problem may be associated with thermal conduction per se.

Thermal conduction has been assumed in theoretical models but not
constrained empirically through observation because interfaces 
at a few $\times10^5$~K require difficult UV observations.
Traditionally interfaces have been studied using spectral
lines of \CIV, \NV, and \OVI; however, these species 
can be photoionized by hot stars (especially
PN central stars with effective temperatures $>$100,000~K) and the
detection of these narrow absorption lines can be hampered by the
stellar P~Cygni profile and confused by interstellar absorption 
lines.  

We have chosen to study the interface layer in the PN NGC\,6543
using \OVI\ emission because the physical properties of its 
interior hot gas have been established by {\it Chandra}
observations \citep{Cetal01} and its central star is too cool to
produce \OVI\ by photoionization 
\citep[$T_{\rm eff} \sim 50,000$~K,][]{Zetal97}.  
{\it Far Ultraviolet Spectroscopic Explorer} ({\it FUSE}) observations
of NGC\,6543 have been made and \OVI\ emission is indeed detected.
In this paper we describe the observations 
and data processing in \S2, present the results in \S3, 
and discuss their implications in \S4.

\section{\fuse\ Observations and Reduction}

We have observed NGC\,6543 with \fuse\ to search for nebular \OVI\ 
emission.
The \fuse\ observatory has four spectrographs operating simultaneously 
to cover the 905--1187 \AA\ wavelength range with high resolution, 
$\lambda/\delta\lambda \simeq $20,000.  Details of the design and 
performance of the \fuse\ spectrographs are described by 
\citet{Metal00} and \citet{Setal00}.  

\begin{figure}[th]
\plotone{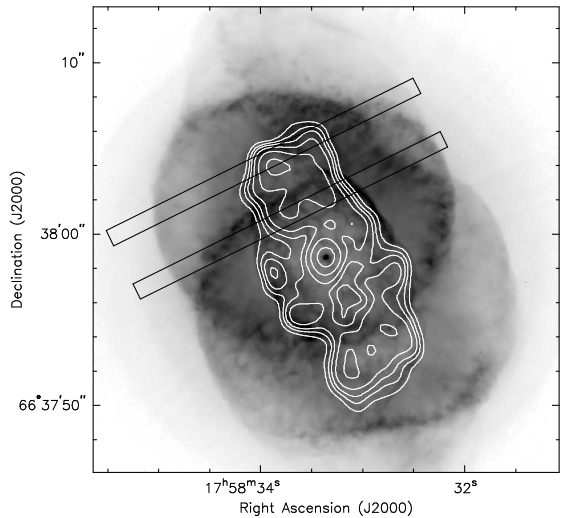}
\caption{{\it Hubble Space Telescope} H$\alpha$ image superposed with 
X-ray contours extracted from a smoothed Chandra ACIS-S image
of NGC\,6543 \citep{Cetal01}. The approximate location of the 
\fuse\ HIRS apertures are marked by rectangles.}
\end{figure}

Three observations of NGC\,6543 were obtained in the time-tag 
mode using the HIRS aperture (1\farcs25$\times$20\arcsec):
``N-Cav'' samples the northern edge of the central cavity,
``N-Ext'' the northern extension of the cavity, and
``OFF'' a region outside the nebula at 150\arcsec\ north of
the central star (see Fig.~1 for their locations).
Additional observations of the central star, available
in the archive and referred to as CS, used the MDRS aperture 
(4\arcsec$\times$20\arcsec) and were acquired in the histogram mode. 

Due to flexure of the instrument, only the LiF1 detector 
segment remains fixed with respect to a position on the sky
throughout an orbit. 
Thus, we only analyzed observations obtained with the LiF1a 
detector segment.
The individual exposures were reprocessed using the \fuse\ 
calibration pipeline software, CALFUSE (version 2.4.0), to extract a 
1-dimensional spectrum for each exposure.  
The observations of the central star obtained in histogram mode were 
also processed using the CALFUSE pipeline with its default settings.

The diagnostics produced by the CALFUSE pipeline were examined to 
ensure that the spectral extraction window was properly centered, 
and to exclude burst/flare events and periods when the central star 
drifted into the HIRS aperture.
Spectral drift between exposures was removed by cross-correlating the 
individual exposures relative to the longest exposure.  The typical 
offset found was generally less than 1 pixel ($\sim$0.0067\AA) and 
never more than 5 pixels.  
The individual exposures for each position were then weighted by 
their exposure times and combined.
The coadded spectra for the N-Cav, N-Ext, OFF, and CS
positions had total effective exposure times of 14.4, 13.7, 37.3, 
and 2.3 ks, respectively.

\section{Results}

The four \fuse\ spectra of NGC\,6543 are presented in Figure 2.
The CS spectrum shows very broad P Cygni profile of the 
stellar \OVI\ lines with numerous interstellar absorption lines of 
H$_2$ and low ionization ions.
The OFF position shows only airglow lines of Ly$\beta$ and \ion{O}{1}.
In contrast, the spectra at the N-Cav and N-Ext show 
prominent narrow emission lines of \ion{C}{2} 
$\lambda\lambda$1036,1037, broad emission lines of \OVI\ 
$\lambda\lambda$1032,1038, and a broad unidentified emission 
line at 1034.432 \AA, superposed on continuum emission that 
has a spectral shape similar to that of the CS.

\begin{figure}
\epsscale{.80}
\plotone{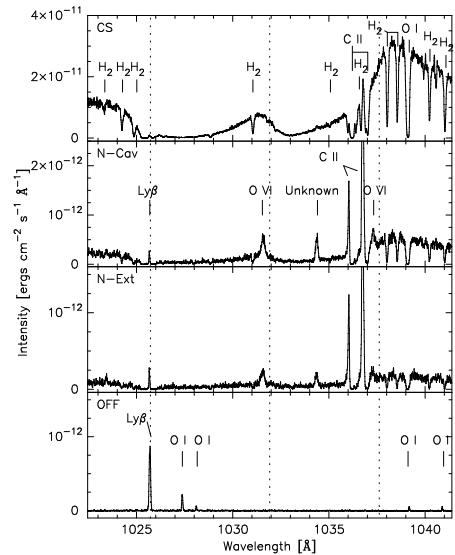}
\caption{\fuse\ spectra in the wavelength range around the \OVI\ 
doublet.  The aperture location is marked in the upper left corner
of each panel.  
The dashed lines mark the rest wavelengths (V$_{\rm hel}=$0 km~s$^{-1}$) 
for Ly$\beta$, and \OVI\ $\lambda\lambda$1032,1038 lines.  
Absorption features of H$_2$, \ion{C}{2} and \ion{O}{1} are identified 
for the CS spectrum, emission features are identified for the N-Cav 
spectrum, and airglow features are marked for the OFF spectrum.}
\end{figure}

The continuum emission in the nebular spectra is most likely
stellar emission scattered by dust in the nebular gas.
To remove this contamination, we have scaled the stellar spectrum 
by factors of 0.018 and 0.0075 and subtracted it from the nebular 
spectra of N-Cav and N-Ext, respectively.
An expanded view of the \OVI\ lines
in Figure 3 shows that the continuum contamination is 
effectively removed and the weaker \OVI\ $\lambda$1038 line 
becomes more prominent.  
We determine the centroid velocity and FWHM of each line by 
fitting a Gaussian profile, but measure the line intensity 
by direct summation.  
The \OVI\ $\lambda$1038 line flux and shape should be regarded 
with some skepticism because the emission line is close to a 
strong \ion{C}{2} absorption line and is in a spectral region where
the P~Cygni profile from the central star may be oversubtracted.
The results of these measurements are summarized in 
Table~\ref{tab_linefit}. 

\begin{figure}
\epsscale{1.0}
\plotone{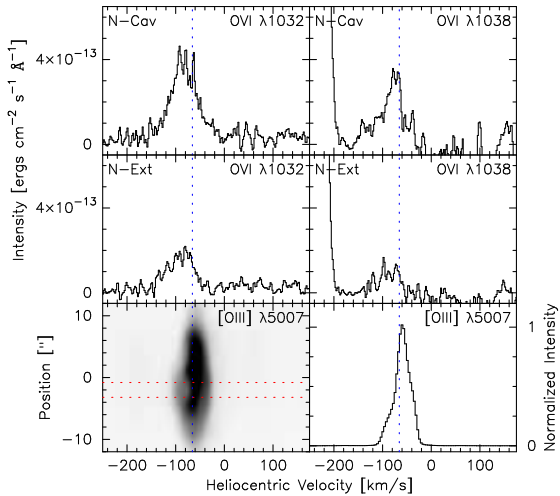}
\caption{Detailed \fuse\ spectra for the \OVI\ $\lambda\lambda$1032,1038 
lines from the N-Cav (top), and the N-Ext (middle).  For comparison the 
bottom panels show \OIII\ $\lambda$5007 nebular emission from a 
long-slit spectroscopic observation 4\farcs5 north of the central 
star. The bottom left panel shows the echellogram and the bottom right
shows the line profile extracted from the position marked by the 
horizontal dashed lines on the echellogram.  The systemic 
velocity of NGC\,6543 is marked by a dashed vertical line in each 
panel.}
\end{figure}

\begin{deluxetable}{llccc}
\tablewidth{0pt}
\tablecaption{Summary of \OVI\ Emission Line Detections\label{tab_linefit}}
\tablehead{
\colhead{} & \colhead{} & \colhead{} & \colhead{} & \colhead{Intrinsic} \\
\colhead{Obs.} & \colhead{Line} & \colhead{Centroid}  & 
\colhead{FWHM} & \colhead{Intensity} \\
\colhead{} & \colhead{} & \colhead{[km~s$^{-1}$]\tablenotemark{a}} & 
\colhead{[km~s$^{-1}$]} & \colhead{[erg~cm$^{-2}$~s$^{-1}$]} 
}
\startdata

N-Cav  & $\lambda$1032 & $-$83.1 & 56.0 & 2.3$^{+0.8}_{-0.9}\times$10$^{-13}$ \\
       & $\lambda$1038 & $-$79.9 & 45.9 & 1.2$^{+0.4}_{-0.5}\times$10$^{-13}$ \\
       &               &         &      &                       \\
N-Ext  & $\lambda$1032 & $-$88.9 & 67.1 & 1.1$^{+0.4}_{-0.4}\times$10$^{-13}$ \\
       & $\lambda$1038 & $-$88.6 & 51.4 & 6.7$^{+2.3}_{-2.1}\times$10$^{-14}$ \\
\enddata

\tablenotetext{a}{All velocities are in the heliocentric frame and assume 
rest wavelength for the \OVI\ transitions of 1031.9261 and 1037.6167 \AA\
\citep{M03}.}

\end{deluxetable}

To convert from the observed to the intrinsic intensity of the \OVI\ 
emission from NGC\,6543, we adopt $E(1034$\AA$ -V) \simeq 10\,E(B-V)$,
or $A_{1034} \simeq 13.2\,E(B-V)$ determined from the parameterized far-UV
extinction curve of \citet{Setal02}.
The optical extinction of NGC\,6543 has been determined by \citet{WL04}
using the Balmer decrement as determined from {\it HST} WFPC2 images 
in the H$\alpha$ and H$\beta$ lines.
It is shown that the extinction is fairly uniform across NGC\,6543
and the average logarithmic extinction at H$\beta$ is 
$c({\rm H}\beta)$ = 0.1.
As $c({\rm H}\beta) = 1.46\,E(B-V)$, the extinction at 1034\AA\
is $A_{1034}$ = 0.90$\pm$0.33, where the uncertainty is based on the
observational scatter seen in far-UV extinction measurements 
\citep{Setal02}.
The intrinsic \OVI\ line intensity is therefore 2.3$^{+0.8}_{-0.9}$ times 
the observed intensity.

In Figure~3 we further compare the velocity profile 
of the \OVI\ lines to that of the \OIII $\lambda$5007 line
from an east-west oriented long-slit echelle spectroscopic observation 
taken at a position 4\farcs5 north of the central star.
The brightest emission component in the \OIII\ line image has
low velocities and corresponds to the dense envelope of the nebular
shell, and the faint expanding ``blister'' near the center of the 
line image corresponds to the northern extension of the central
cavity.  The heliocentric velocity of the \OVI\ emission is similar 
to that of the expanding blister in the \OIII\ line indicating
that they are physically associated.

\section{Discussion}

\fuse\ observations of NGC\,6543 and NGC\,7009 \citep{Ietal02} 
detect, for the first time, \OVI\ emission from PNe.  
Since the central star of NGC\,6543 is too cool to photoionize
\OV, the \OVI\ ions must be produced by thermal collisions
in gas at temperatures of a few $\times10^5$~K at the 
interface layer between the cool nebular shell and the 
hot gas in the PN interior.  
Therefore the intensity of \OVI\
in NGC\,6543 can be used to investigate the physical processes
at the interface.

Are the \fuse\ observations of \ion{O}{6} from NGC\,6543 consistent
with that expected from thermal conduction in such an interface layer?
To model the expected \ion{O}{6} emission, we assume that the
PN shell is a pressure-driven bubble and the radiative losses of the 
hot interior gas have a negligible effect on the dynamics.
We further assume a constant AGB mass loss, resulting in a radial
density profile $\propto radius^{-2}$ for the circumstellar medium.
The dynamics of such a bubble can be solved analytically 
\citep[e.g.,][]{GM95} and is characterized by a pressure and age which 
can be expressed in terms of observed quantities such as the
fast stellar wind terminal velocity and mass loss rate,
and the nebular shell size and expansion velocity.
To calculate the temperature and density structure at 
the interface layer, radiative losses and thermal conduction 
are considered by numerically solving the continuity and energy 
equations \citep[i.e., Equations (42) and (43) of ][]{Wetal77} which 
rely directly upon the pressure and age found from the analytic solution.



The central star of NGC\,6543 has a fast wind terminal velocity
of 1750 km~s$^{-1}$ and a mass loss rate of 4.0$\times$10$^{-8}$ 
M$_\odot$~yr$^{-1}$ \citep{PCSL89}.
Adopting a distance of 1.0$\pm$0.3 kpc \citep{Retal99}, 
the radii of N-Cav and N-Ext are 0.02 and 0.04 pc, respectively.  
The expansion velocity of NGC\,6543's nebular shell near the nebular
minor axis is $\sim$20 km~s$^{-1}$ \citep{MS92}, which is appropriate 
for the N-Cav position.  The [\ion{O}{3}] emission associated with 
the N-Ext shows a velocity offset of 
$\sim$30 km~s$^{-1}$ from the systemic velocity of NGC\,6543 
(see Fig.~3); for an inclination angle of 35$^\circ$ \citep{MS92}
the expansion velocity will be $\sim$40 km~s$^{-1}$.

Use of these observational parameters results in an analytic solution
with a nebular age of $\sim$1000 yrs and a pressure in the central 
cavity of $\sim$3$\times$10$^{-7}$ dyne~cm$^{-2}$.  If we compare
these values with independent estimates we find that the age is 
reasonable for NGC\,6543 \citep[e.g.,][]{Retal99} but the pressure
is roughly a factor of 10 higher than indicated by observations of
both the cool nebular shell and the hot gas interior.  Consequently
this simple model would over-produce $L_{\rm X}$ as did the models 
of \citet{ZP98}.  Likewise, using the method outlined later, we
find the expected \ion{O}{6} emission to be $\sim$10 times higher than 
observed.  Both of these discrepencies arise from the incorrect 
model prediction of pressure within the central cavity, and are likely
caused by incorrect assumptions about the fast and AGB wind mass loss 
rates.

To obtain an alternative estimate of the \ion{O}{6} emission we
consider the observed pressure in the nebular shell and hot interior.
In the nebular shell \citet{Zetal04} find an electron density of 
$\sim$6300~cm$^{-3}$ and temperature of 6800~K, and \citet{RTG04}
find $\sim$10000~cm$^{-3}$ and $\sim$8000~K, indicating a pressure 
of 1.2--2.2$\times$10$^{-8}$ dyne~cm$^{-2}$.  For the interior hot
gas \citet{Cetal01} find that the X-ray emitting gas has a plasma 
temperature of $\sim$1.6$\times$10$^6$~K and electron density of 
$\sim$50$\epsilon^{-1/2}$, where $\epsilon$ is the filling factor.
For $\epsilon=$0.5, the pressure in the central cavity is 
$\sim$3$\times$10$^{-8}$ dyne~cm$^{-2}$, similar to the pressure in the
nebular shell.
We adopt the pressure obtained for the hot gas 
in the continuity and energy equations to determine the 
temperature and density structure of the interface layer.  
The resulting temperature and density profile applicable 
to the N-Cav observations are plotted as a function of distance 
from the inner wall of the nebular shell in Figure~4.

\begin{figure}
\epsscale{.80}
\plotone{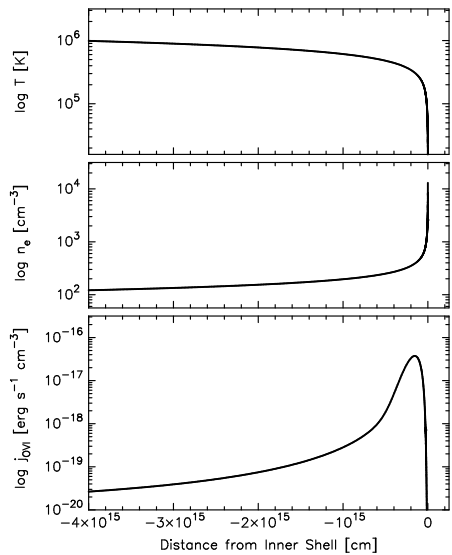}
\caption{Results of the model calculation for the N-Cav observation 
of NGC\,6543 showing profiles for the temperature (top), electron 
density (middle), \OVI\ emissivity (bottom) as a function of 
distance from the inner wall of the nebular shell.}
\end{figure}

The \OVI\ density can be determined from the temperature and density
profile by assuming ionization equilibrium and adopting a nebular
oxygen abundance relative to hydrogen of 5.6$\times$10$^{-4}$ 
\citep{AC83,BSetal03}.  The \OVI\ $\lambda$ 1032,1038 line intensity, 
$I_{\rm O\,VI}$, in ergs~s$^{-1}$~cm$^{-2}$ can be derived from 
equations 4.4 and 4.11 of \citet{S78} and expressed as:
\begin{equation}
I_{\rm O\,VI} = n_{\rm e}~n_{{\rm O\,VI}}~h\nu~
{{8.629\times10^{-6}}\over{T^{1/2}}} 
{{\bar{\Omega}_{{\rm O\,VI}}(T)}\over{g_j}}~
e^{-\chi / kT } {V\over{4 \pi d^2}} 
\end{equation}
where $n_{\rm e}$ and $n_{\rm O\,VI}$ are respectively the  
number densities of electrons and \OVI\ ions in units of
cm$^{-3}$, $h\nu$ is the average photon energy in ergs, 
$T$ is the temperature in Kelvin, $\bar{\Omega}_{{\rm O\,VI}}(T)$ 
is the Maxwellian-averaged collision strength\footnote{
A parameterization for $\bar{\Omega}_{{\rm O\,VI}}(T)$
is given as equation (6) of \citet{SS94}.}, 
$g_j$ is the statistical weight of the lower energy state 
(in this case $g_j=2$), $\chi$ is the excitation energy of the upper 
energy state (in this case $\chi=h\nu=1.92\times10^{-11}$~ergs), 
$V$ is the emitting volume in cm$^3$, and $d$ is the distance to the 
object in cm.

The \OVI\ emission per unit volume, $j_{\rm O\,VI}$, is plotted in 
Figure~4 and suggests the thickness of the emitting layer would be
$\sim$5$\times$10$^{14}$~cm (0.00016~pc).
To simulate the \fuse\ observations of N-Cav and N-Ext 
we integrate $j_{\rm O\,VI}$ over a short cylinder corresponding to 
the volume intersected by the HIRS aperture, i.e., 4\arcsec-radius 
and 1\farcs25-height).  Based on these calculations we would 
expect 5.2$\times$10$^{-13}$ erg~cm$^{-2}$~s$^{-1}$ 
and 2.6$\times$10$^{-13}$ erg~cm$^{-2}$~s$^{-1}$ at the N-Cav and 
N-Ext positions, respectively.  
While the predicted \OVI\ emission intensity is roughly 1.5 
times higher than observed, the two are consistent given the
uncertainties in the model parameters.

In summary, we have detected nebular \OVI\ emission from NGC\,6543
from hot gas that is most likely at the interface between the 
nebular shell and the hot, X-ray-emitting gas within the PN interior.
We have used the fast wind and nebular properties to model the \ion{O}{6}
emission but find large discrepancies similar to those found for
$L_{\rm X}$.  This may result from overly simplistic assumptions
for the fast and AGB wind mass loss rates.  The observed pressures in the
nebular shell and interior hot gas are similar.  Adopting the hot gas
pressure for the interface layer, we find that the model with 
thermal conduction predicts \OVI\ emission consistent with the 
observations.

\acknowledgements

This work was supported through the \fuse\ grant NASA~NAG~5-12255.
We would like to thank the anonymous referee for his/her insightful
comments which have greatly improved this work.
We thank B.\ Boroson for sharing his code to model the ionization
structure of a conduction layer in a wind-blown bubble.  
M.A.G.\ acknowledges the support from the grant AYA 2002-00376 of
the Spanish MCyT (cofunded by FEDER funds).

\end{document}